\documentclass[%
 reprint,
 amsmath,amssymb,
 aps,
]{revtex4-1}

\usepackage[margin=1.0in]{geometry}

\usepackage[english]{babel}
\usepackage{graphicx}
\usepackage[]{xcolor}

\usepackage{epstopdf, epsfig}
\usepackage{amsmath}
\usepackage{amsmath,amsfonts,amssymb}
\usepackage{amssymb}

\usepackage{soul}
\usepackage{mathrsfs} 

\usepackage{tikz}
\usepackage{graphicx,wrapfig,lipsum}
\usepackage{enumitem}
\usepackage{paralist}
\usepackage[belowskip=-5pt,aboveskip=0pt]{caption}

\usepackage[compact]{titlesec}
\titlespacing{\section}{0pt}{2ex}{0ex}
\titlespacing{\subsection}{2pt}{1ex}{2ex}
\titlespacing{\subsubsection}{0pt}{0.5ex}{0ex}

\usepackage{cleveref}
\crefformat{section}{\S#2#1#3} 
\crefformat{subsection}{\S#2#1#3}
\crefformat{subsubsection}{\S#2#1#3}

\usepackage{wrapfig}

\usepackage{bm}
\usepackage{algorithm}

\begin{document}

\preprint{APS/123-QED}

\title{Turbulence modulation by suspended finite-sized particles - Towards physics-based multiphase subgrid modeling}
\thanks{Supplementary materials can be downloaded from https://github.com/jasonshen1990/Physics-based-multiphase-subgrid-modeling}%

\author{S. Balachandar}
\email{bala1s@ufl.edu}
\affiliation{%
 University of Florida, Gainesville, FL, 32608, USA
}%


\author{C. Peng}
\email{pengcheng@sdu.edu.cn}
\affiliation{
 Shandong University, Jinan, Shandong, China
}%

\author{L.-P. Wang}
\email{wanglp@sustech.edu.cn}
\affiliation{%
 Soutern University of Science and Technology, Shenzhen, Guangdong, China
}%


\date{\today}

\begin{abstract}
The presence of a dispersed phase substantially modifies small-scale turbulence. However, there has not been a comprehensive mechanistically-based understanding to predict turbulence modulation. Based on the energy flux balance, we propose a theoretical model  to predict the turbulent kinetic energy modulation in isotropic turbulence due to the dispersed phase. The comparison between model predictions and results from particle-resolved simulations and high-fidelity experiments validates the performance of the model over a wide range of turbulence and particle parameters. 
\end{abstract}

\maketitle


\section{Introduction}

The presence of particles, droplets, or bubbles in a flow substantially alters the nature of multiphase turbulence, rendering the problem far more complex than single-phase turbulence. 
In homogeneous isotropic turbulence laden with particles of negligible sedimentation, a pivot scale of the order of the particle diameter ($D$) is found to distinguish whether turbulence is attenuated or augmented 
\cite{ten2004fully, lucci2010modulation, yeo2010modulation, gao2013lattice, oka2022attenuation, peng2023parameterization}.
In inhomogeneous and wall-bounded flows with sedimenting particles, 
turbulence modulation is more complex. In some cases, the entire turbulence is due to the suspended particles, without a pivot scale  \cite{cantero2009turbidity, shringarpure2012dynamics}.

Several criteria for turbulence modulation have been advanced in the past. Gore \& Crowe \cite{gore1989effect} suggested that turbulence is augmented if the ratio of $D$ to the characteristic size of the energy-containing eddies is greater than 0.1, and otherwise suppressed. On the other hand, Elghobashi \& Truesdell \cite{elghobashi1993two} observed turbulence enhancement even for particles of diameter comparable to the Kolmogorov scale ($\eta$). 
The Gore \& Crowe criterion was recently updated by Oka \& Goto \cite{oka2022attenuation} by requiring $D$ to be not only below the integral length scale but also larger than the Taylor microscale divided by the square root of particle-to-fluid density ratio for turbulence attenuation.
Hetsroni \cite{hetsroni1989particles} recommended particle Reynolds number $Re_p > 400$ as the criterion for turbulence enhancement resulting from vortex shedding. Bagchi \& Balachandar \cite{bagchi2004response}, however, observed vortex shedding to initiate at much lower $Re_p$ in the presence of free-stream turbulence. 
Tanaka \& Eaton \cite{tanaka2008classification} introduced a particle momentum number as the nondimensional parameter to distinguish between turbulence augmentation and attenuation. A similar criterion has also been introduced by Luo, Luo \& Fan \cite{luo2016turbulence}. Peng et al. \cite{peng2023parameterization} presented empirical correlations that well predicts multiphase turbulence modulation in the absence of gravitational effects.

The purpose of this work is to develop a physics-based closure model for subgrid turbulence that can be used in multiphase large eddy and Reynolds averaged Navier-Stokes simulations (LES \& RANS). 
We focus on the homogeneous isotropic flow configuration, but desire the closure model to be universal with applicability for a wide range of particle sizes (including $D \gg \eta$), volume and mass fractions. Furthermore, we want the model to account for the inertial and gravitational effects on the particles as well as the dissipative effect of inter-particle collisions at higher volume fractions. 

The mesoscale state of the dispersed multiphase flow is considered to be known, {\it e.g.}, as in LES, and we limit the quest to modeling of turbulence modulation at the micro or subgrid scales. Such understanding of turbulence modulation along with well-developed closure models of single-phase turbulence may provide robust and general multiphase subgrid closures.
The modeling of subgrid turbulence  however remains formidable as
a very wide range of scales and a large number of particles are involved. 

Conceptually, we distinguish two different mechanisms of turbulence modulation. At the microscale, the slip velocity between the particles and the fluid due to particle inertia, finite size, and gravity results in pseudo turbulence, altering the spectral distribution of kinetic energy. 
At the mesoscale, turbulence may be modulated by the gravitational influence on a nonuniform distribution of particulates. Buoyancy-induced instabilities enhance turbulence, while stable stratification can strongly suppress turbulence \cite{cantero2009direct, salinas2021anatomy}. 
By limiting attention to only turbulence modulation at the subgrid scale, we avoid the influence of mesoscale turbulence modulation. Furthermore,
we shall assume the particulate phase to be uniformly distributed in the theoretical analysis.
We present a physics-based model to predict turbulence modulation and test it against particle-resolved (PR) simulation and experimental results for isotropic turbulence \cite{ten2004fully, yeo2010modulation, chouippe2019influence, oka2022attenuation, shen2022turbulence, peng2023parameterization, hwang2006homogeneous, hwang2006turbulence} and the central region of turbulent channel flow \cite{peng2019direct, shen2023turbulence}. 
The validated model is then used to illustrate turbulence modulation over a wider parameter space.

\section{Theoretical Model}
Consider an Euler-Euler (EE) LES of particle-laden flow with a random distribution of particles 
in a finite-volume cell of size $\Delta x \gg D, \eta$. 
Let the mean fluid velocity  $\bf{u}$, particle velocity $\bf{v}$, and particle volume fraction $\phi$ be known within the cell. 
{From the energy transfer of the resolved-scale turbulence, we estimate the flux of kinetic energy to the subgrid scales, which is taken to be equal to the average viscous dissipation rate $\epsilon$ in the bulk of the fluid.}
{\it{Then, the multiphase LES subgrid modeling quest is to predict closure quantities such as (i) the subgrid fluid Reynolds stress, (ii) particle Reynolds stress, and (iii) mean and rms force 
acting on the particles 
}} \cite{bala-book2023}. Here, the focus will be on quantifying turbulence modulation in terms of the ratio between multi and single-phase subgrid fluid Reynolds stress.




In the isotropic limit, the four key controlling parameters are \cite{peng2023parameterization} (i) $D/\eta$, (ii) subgrid turbulence intensity measured in terms $Re_\Delta = \epsilon^{1/3} (\Delta x)^{4/3} / \nu$, (iii) particle-to-fluid density ratio $\rho = \rho_p/\rho_f$, and (iv) $\phi$. 
In the presence of a mean relative velocity (i.e., $\bf{u} \ne \bf{v}$), subgrid Reynolds stress tensor is axisymmetric. 
There is an additional parameter: (v) relative mean slip velocity, $u_r = |{\bf{u}}-{\bf{v}}| /u_{k}$, where the denominator is the Kolmogorov velocity.
We propose the following energy flux balance within the subgrid \cite{oka2022attenuation, chouippe2015forcing, chouippe2019influence}
\begin{equation} \label{eq:mp-balance}
\begin{split}
   & \epsilon + {N} \, 3\pi \nu D \, {{\Phi}} |{\bf{u}} - {\bf{v}}|^2 = C_{c,mp} \dfrac{k_{f,mp}^{3/2}}{\Delta x} \\
   &+ C_p {N} \, 3\pi \nu D \, {{\Phi '}} \, \Delta u^2 + C_{co} \dfrac{\rho \phi^2}{D} \Delta u \, k_p\, ,
    \end{split}
\end{equation}
where the second term on the left-hand side is {the rate of work input on the subgrid fluid-particle system due to mean relative motion of all the $N=\phi/(\pi D^3/6)$ particles.} This term contributes to the subgrid energy transfer 
in addition to that from cascading turbulence represented by $\epsilon$. $\Phi(Re,\phi)$ represents correction to Stokes drag due to finite value of $Re = |{\bf{u}} - {\bf{v}}| D/\nu$ and volume fraction $\phi$ (see \cite{richardson1954sedimentation, gidaspow1994multiphase, tenneti2011drag}). 
This term, along with two additional contributions arising from particle acceleration and inter-particle collision, was rigorously derived in \cite{chouippe2015forcing, chouippe2019influence}. The other two contributions are generally small.

The first term on the right-hand side represents fluid phase dissipation in the bulk, where $k_{f,mp}$ is the sub-grid fluid kinetic energy in the multiphase system. 
The second term is dissipation in the immediate neighborhood of the particles that do not contribute to the bulk fluid 
turbulence. The local dissipation depends on the fluctuating relative velocity $\Delta u$ between the particle and the surrounding fluid, which again can be taken to depend on the parameters listed above. In this term, $\Phi'$ is correction to Stokes drag based on $Re' = \Delta u D/\nu = (\Delta u/u_k)(D/\eta)$. From $\Phi'(Re',\phi) \propto C_D Re'$ ($C_D$ is the drag coefficient), we obtain this term to be $\propto \Delta u^3$, in agreement with \cite{oka2022attenuation}.
The third term accounts for the dissipative effect of inter-particle collisions, where $k_p$ is subgrid particle kinetic energy (see supplementary material). This term is expected to play a role only when $\phi \gtrapprox 10$\%. The empirical coefficients $C_{c,mp}$, $C_p$, and $C_{co}$ will be determined by fitting the available experimental and PR simulation data.


Given $\epsilon$, we calculate Kolmogorov length, time, and velocity scales as: $\eta = \nu^{3/4}/\epsilon^{1/4}$, $\tau_{k} = \eta^{2/3}/\epsilon^{1/3}$, and $u_{k}=(\epsilon\eta)^{1/3}$.  The scaling relation for slip velocity by Balachandar \cite{balachandar2009scaling, ling2013scaling} can be restated as
\begin{equation} \label{eq:delu}
    \dfrac{\Delta u}{u_{k}} = \begin{cases}
    \vert 1-\beta \vert St_k  & \mbox{(i) if }{\tau_k} > {\tau_{p}} \\
  \vert 1-\beta \vert St_k^{1/2} & \mbox{(ii) if  } {\tau_k} < {\tau_{p}} < \tau_{\Delta} \\
  \vert 1-\beta\vert Re_\Delta^{1/4}  & \mbox{(iii) if } {\tau_p} > \tau_{\Delta} \\
  u_r & \mbox{(iv) if } u_r \mbox{ dominates}
   ,
\end{cases}
\end{equation}
where  $\beta = 3/(2\rho + 1)$,
particle time scale $\tau_p = (2\rho +1)D^2/(36 \nu \Phi')$, 
$St_k = \tau_p/\tau_k$ is the particle Stokes number based on the Kolmogorov time scale, and $\tau_{\Delta} = (\Delta x)^{2/3} /\epsilon^{1/3}$.
The four regimes are (i) small, (ii) medium, (iii) large, and (iv) rapidly settling particles. 
We note that $\tau_p/\tau_k = (2\rho +1) (D/\eta)^2/(36 \Phi')$ and$\tau_p/\tau_\Delta = (\tau_p/\tau_k)/\sqrt{Re_\Delta}$.
A simplified evaluation of the implicit equation, Eq.~(\ref{eq:delu}), is discussed in \cite{balachandar2009scaling}.
Eq.~(\ref{eq:delu}) was obtained in the absence of two-way coupling. With the effect of turbulence modulation, the estimated slip velocity must be adjusted by {dividing} $\Delta u/u_{k}$ given in \eqref{eq:delu} by $\sqrt{k_{f,mp}/k_{f,sp}}$.

In order to evaluate turbulence modulation as the ratio, $k_{f,mp}/k_{f,sp}$, between multi and single-phase kinetic energy, for the same energy flux $\epsilon$, we first define the the single-phase limit as
$\epsilon = C_{c,sp} {k_{f,sp}^{3/2}}/{\Delta x}$, which is {similar to the first term on the right-hand sides of} \eqref{eq:mp-balance}. We divide \eqref{eq:mp-balance} by the above single-phase $\epsilon$ to obtain
\begin{equation} \label{eq:krat}
\begin{array} {ll}
     &\left( C' + C_{co} \rho \phi^2 Re_\Delta^{\text{\tiny{1/2}}}  \frac{\Delta u}{u_{k,sp}} \frac{k_p}{k_{f,mp}} \frac{\eta_{sp}}{D}  \right) \left( \frac{k_{f,mp}}{k_{f,sp}} \right)^{\text{\tiny{3/2}}} \\
     &+ 18 C_p \phi \, \Phi' \, \left( \frac{\Delta u}{u_{k,sp}} \frac{\eta_{sp}}{D} \right)^{\text{\tiny{2}}} \frac{k_{f,mp}}{k_{f,sp}} \\
     & = 1 + 18 \phi \, \Phi \, u_r^{\text{\tiny{2}}} \left(\frac{\eta_{sp}}{D} \right)^{\text{\tiny{2}}}   \, ,
\end{array}
\end{equation}
where $C' = {C_{c,mp}}/{C_{c,sp}}$
is another empirical coefficient that must be determined.
The above is an implicit equation for the ratio ${k_{f,mp}}/{k_{f,sp}}$ in terms of the five input parameters (note ${\Delta u}/{u_{k,sp}}$ is a function of the five parameters). In the limit of significant dissipation due to inter-particle collisions, particle-to-fluid subgrid kinetic energy ratio, ${k_p}/{k_{f,mp}}$, must also be specified, whose modeling can follow the work of Wang and Stock \cite{wang1993dispersion}.

\section{Evaluation of Theory}

We now validate the model by reproducing results on turbulence modulation from past PR simulations and experiments. In obtaining \eqref{eq:krat} it has been taken that the energy flux $\epsilon$ into the subgrid scales is the same for both single and multiphase cases. While this is appropriate for LES closure, in the forced isotropic conditions of the simulations to be compared, the forcing at the largest scales is maintained the same, which does not guarantee the dissipation rates of single and multiphase turbulence to be the same. 
We have also ignored the effect of inter-particle collisions.
Given the five non-dimensional parameters,
the above equation can be solved for the ratio ${k_{f,mp}}/{k_{f,sp}}$, with the additional information on the dissipation ratio ${\epsilon_{mp}}/{\epsilon_{sp}}$. 
We have replaced $Re_\Delta$ with the Taylor microscale Reynolds number, but the dependence on $Re_\Delta$ is quite weak 

\begin{figure}[htb]
    \includegraphics[width=60mm] {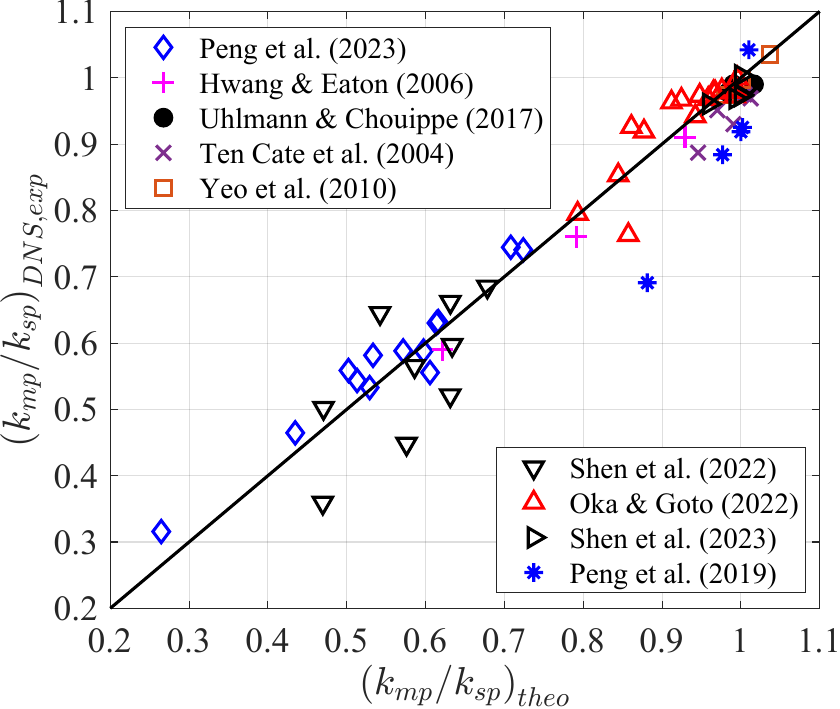}
	\caption{Comparison of turbulence modulation obtained in simulations/experiments (y-axis) against theoretical prediction using \eqref{eq:krat} (x-axis). 
 }
	\label{fig1}
\end{figure} 

We consider 49 PR simulations from six different sources \cite{ten2004fully, yeo2010modulation, chouippe2019influence, shen2022turbulence, peng2023parameterization}
and 3 experimental data from Hwang \& Eaton \cite{hwang2006homogeneous, hwang2006turbulence}. They cover: $D/\eta \in [0.96, 17.77]$, $Re_\lambda \in [32.95, 240]$, $\rho \in [0, 2080]$, and $\phi \in [7.17 \times 10^{-6}, 0.12]$. Particle settling is negligible and therefore $u_r = 0$. 
We observe good agreement for $C_p = 1.0$ and 
\begin{equation} \label{eq:C'}
\begin{split}
    C' -1 = \min\{(\rho-1)\phi, 0.48\} \quad \quad \quad \\ \left( 1 - \sigma(\ln(St_k) - \ln(500))) \right) \, ,
\end{split}
\end{equation}
where $\sigma$ is the sigmoid function. 
Figure \ref{fig1} presents the actual measured value of turbulence modulation $({k_{f,mp}}/{k_{f,sp}})_{dns,exp}$ plotted again that predicted by theory. We observe the agreement to be quite good.
As observed by prior researchers, in the absence of gravitational effect, turbulence is generally attenuated and the attenuation can be substantial.

For heavy particles, the coefficient $C'$ is larger than unity and in \eqref{eq:C'} the difference is expressed as two parts, one that depends on excess mass loading by the particles and the other dependent on particle Stokes number.
The first factor is motivated by Peng {\it et al.}~\cite{peng2023parameterization}, who observed increased mass of the multiphase flow to be an important parameter. The rationale is that with increasing mixture density, the fluid velocity fluctuation decreases. However, with increased mass loading particles become less responsive, and the fluid velocity fluctuation is less influenced by the particles. Therefore, here we find it is necessary to cap the value of $C'$. The Stokes number-dependent second factor is motivated by  the observation in \cite{oka2022attenuation} that when $St_k$ increases above a few hundred, the attenuation effect decreases. 

Further comparisons are made using the central region of PR turbulent channel flow data. Validation against 10 simulation cases taken from \cite{peng2019direct, shen2023turbulence} are presented in Figure \ref{fig1}. 
Even in the absence of gravitational effect, the average streamwise fluid and particle velocities are different. 
However, in all the 10 cases considered, the effect of mean slip velocity is relatively small. The results presented are observed to be not sensitive to the precise fit used for $C'$.


\section{Parameteric Effect and Comparison}
\begin{figure*}[ht!]
  	\centering
   \includegraphics[width=0.9\linewidth]{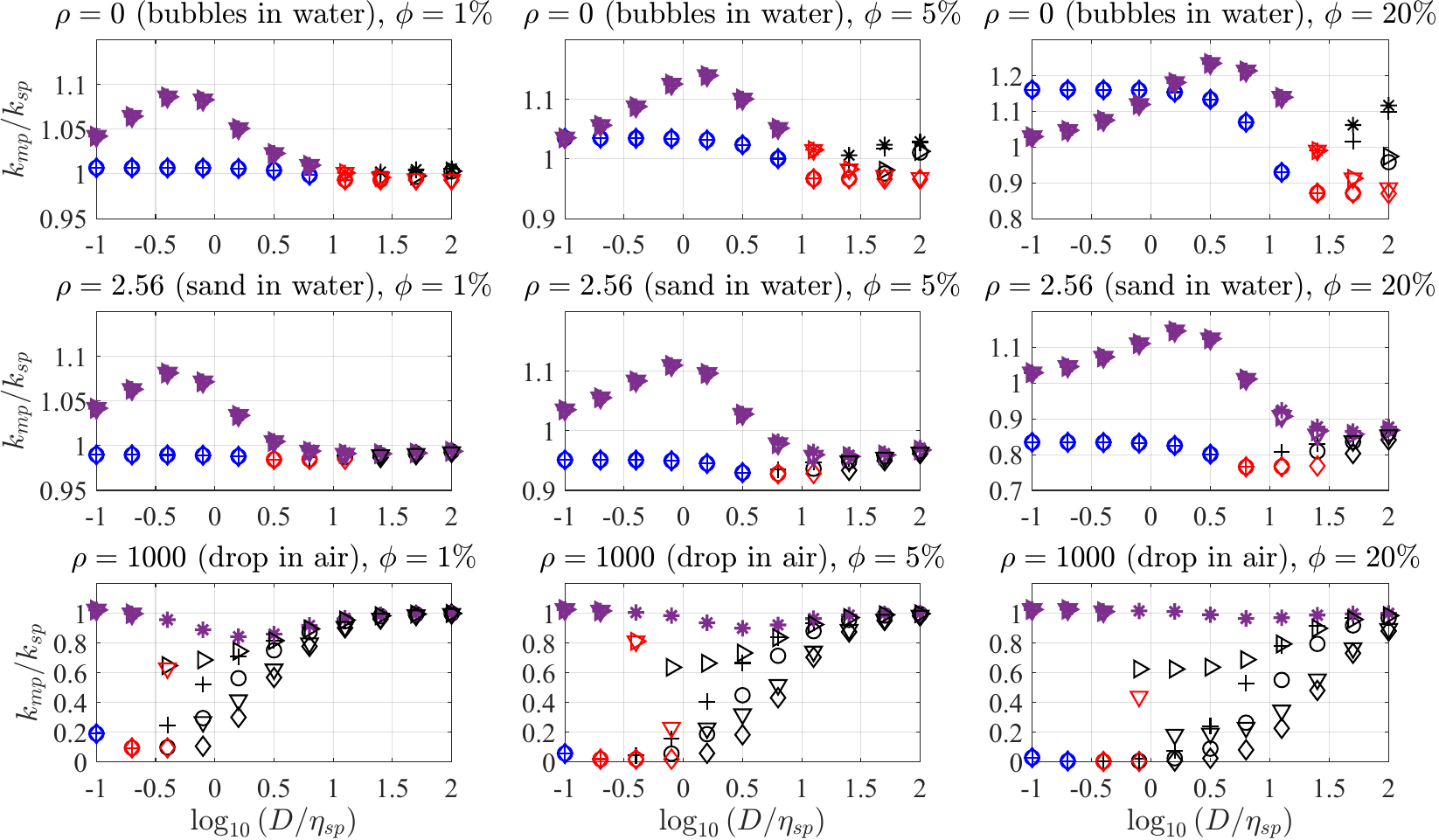}
	\caption{The ratio of multi to single-phase kinetic energy as a function of $D/\eta$: {$(Re_\Delta,u_r) = (5,0)$ (cross), $(40,0)$ (circle), $(400,0)$ (diamond), $(5,2)$ (asterisk), $(40,2)$ (horizontal triangle) and $(400,2)$ (vertical triangle). The blue, red, black, and purple color of the symbols indicate the condition falls into Regime (i), (ii), (iii), and (iv) of $\Delta u/u_{k,sp}$, respectively.}}
	\label{fig2}
\end{figure*} 

In this section, using the model, we investigate the effects of 
$D/\eta$, $Re_\Delta$, $\rho$, $\phi$, and $u_r$. The results for $u_r = 0.0$ and 2.0 are presented in Figure \ref{fig2}. The left, middle, and right columns correspond to volume fractions of 1\%, 5\%, and 20\%. The top, middle, and bottom rows correspond to density ratios of 0 (bubbles in water), 2.56 (sand particles in water), and 1000 (water droplets in air). The six curves correspond to $Re_\Delta = 5$, 40, and 400 for the two values of $u_r$. In all these calculations we have taken $\epsilon_{mp}/\epsilon_{sp} =1$. Each symbol is colored according to its slip velocity regime given in \eqref{eq:delu}. {\it{A Matlab code is provided as supplementary material.}}

First, we consider $u_r = 0$ limit of weak settling.
In case of small volume fraction of very light particles (bubbles), the turbulence modulation effect is quite small. However, augmentation and attenuation are observed for particles of size smaller and larger than about $10\eta_{sp}$. 
The augmentation is mainly due to the reduced effective density of the mixture and at larger sizes, the relative velocity increases, and the associated dissipation around the particles contributes to effective turbulence attenuation. 


For the density ratio of 2.56, there is no turbulence augmentation, and attenuation is maximized at intermediate particle sizes of $1 < D/\eta_{sp} < 10 $. The effect of $Re_\Delta$ is not strong. In the case of small particles of $\rho = 1000$, the substantial damping is due to the increase in the mixture density. Small particles tend to move with the fluid and with increased mixture density, for the same energy flux, the intensity of turbulence decreases. In contrast, larger particles remain relatively stationary with their subgrid velocity fluctuations being much smaller than that of the fluid, attenuation is mostly due to dissipation associated with the relative velocity. 


With increasing $u_r$, $\Delta u$ is dictated more by $u_r$, than by cascading turbulence estimated in \eqref{eq:delu}. 
The large relative velocity contributes to additional subgrid rate of work and as a result there is turbulence augmentation in all cases considered. For $u_r = 2$, augmentation effect reaches a peak at around $D/\eta_{sp} \sim 0.5$ in a dilute system. With increasing volume fraction, the amplitude of peak turbulence augmentation increases and the location shifts to $D/\eta_{sp} \sim 3$. In interpreting the results for large $u_r$ it must be noted that such large slip velocity either by gravitational settling or inertial response to larger resolved-scale eddies is generally associated with larger values of $D/\eta$.


With the relation $\Delta x/\eta = Re_\Delta^{3/4}$, the Gore \& Crowe criterion can be rewritten as $D/\eta > 0.1 Re_\Delta^{3/4}$. 
Non-dimensional settling velocity can be expressed as
${V_s}/{u_k} = \left({D}/{\eta}\right)^2 {|\rho-1| \, g \, \eta^3}/{(18 \nu^2 \, \Phi)} $.
Now, if we take $\eta \sim 100 \, \mu$m, then for water droplets in air we obtain $V_s/u_k \sim 10 (D/\eta)^2$, and for sand particles or bubbles in water we obtain $V_s/u_k \sim (D/\eta)^2$. In general, it can be concluded that larger particle sizes correspond to $u_r \gtrapprox 10$. Thus, turbulence enhancement for larger particles is due to production resulting from large settling-induced relative velocity. 


At $u_r = 1$, in all cases, $Re_p$ approaches a value of $\approx 100$ for $D/\eta > 30$ and $Re_p$ increases (decreases) with increasing (decreasing) $u_r$. For example, at $u_r = 10$, $Re_p \approx 100$ for $D/\eta > 10$. Thus, Hetsroni's criterion for turbulence augmentation can be reinterpreted as a requirement for turbulence production due to large relative velocity. 
Our results however show that turbulence augmentation can occur even in the case of smaller Kolmogorov-scale particles of small relative velocity, provided $\rho$ is not large.


Luo~{\it et al.}~\cite{luo2016turbulence}  predict augmentation when $\rho (D/H)^{-1}Re_b^{-11/16}Re_p > 7000$, while Yu et al.'s criterion is  $Re_p \phi^{0.1}Re_b^{-0.53}(D/H)^{-0.61}\rho^{-0.065} > 1.55$, where $Re_b$ and $H$ are the bulk Reynolds number and half channel width of the turbulent channel flow, respectively. Again, the significant dependency on $Re_p$ in these criteria can be interpreted as the requirement for sufficient $u_r$ to trigger turbulence enhancement. Both models indicate turbulence augmentation can happen at smaller $Re_p$ when $D$ and $Re_b$ decrease. 
However, the current model predicts non-monotonic dependence of turbulence enhancement on $D$. 
For small particles, turbulence enhancement becomes weaker as the particle size decreases.  


\section{Conclusions}
A simple model of turbulence modulation induced by particles/droplets/bubbles is proposed based on an energy flux balance within a representative volume. 
The size of the representative volume is assumed to fall within the inertial subrange, and 
be larger than 
the particle diameter. 
The energy flux balance considers the work input due to the interphase mean slip and added viscous dissipation occurring at the particle-fluid interfaces due to the relative fluctuating motion.
This balance brings in the effects of all important parameters of the system, namely, the particle size, volume fraction, density ratio, mean slip velocity, and the representative volume-scale $Re$.
This model represents one of the first efforts to mechanistically quantify the feedback effects of the
dispersed phase on the subgrid Reynolds stress of fluid turbulence in a multiphase large eddy simulation, so a coarse-grained simulation can be reliably conducted.
 
The model predictions of turbulence modulation
agrees well with particle-resolved simulations and experimental results. 
Using this model, we explored the roles of each parameter on turbulence modulation.
Both attenuation and augmentation of turbulence are found with light and heavy particles. 
With the proposed model, we hope to illustrate not only how the governing parameters affect turbulence modulation, but also point to a physically meaningful way to gather and organize future simulations and experiments on turbulence modulation. As more data becomes available, the model should be refined and extended.



\appendix

\bibliography{references}

\begin{thebibliography}{10}

\bibitem{boivin1998direct}
Marc Boivin, Olivier Simonin, and Kyle~D Squires.
\newblock Direct numerical simulation of turbulence modulation by particles in
  isotropic turbulence.
\newblock {\em Journal of Fluid Mechanics}, 375:235--263, 1998.

\bibitem{ten2004fully}
Andreas Ten~Cate, Jos~J Derksen, Luis~M Portela, and Harry~EA Van Den~Akker.
\newblock Fully resolved simulations of colliding monodisperse spheres in
  forced isotropic turbulence.
\newblock {\em Journal of Fluid Mechanics}, 519:233--271, 2004.

\bibitem{lucci2010modulation}
Francesco Lucci, Antonino Ferrante, and Said Elghobashi.
\newblock Modulation of isotropic turbulence by particles of taylor
  length-scale size.
\newblock {\em Journal of Fluid Mechanics}, 650:5--55, 2010.

\bibitem{yeo2010modulation}
Kyongmin Yeo, Suchuan Dong, Eric Climent, and Martin~R Maxey.
\newblock Modulation of homogeneous turbulence seeded with finite size bubbles
  or particles.
\newblock {\em International Journal of Multiphase Flow}, 36(3):221--233, 2010.

\bibitem{gao2013lattice}
Hui Gao, Hui Li, and Lian-Ping Wang.
\newblock Lattice boltzmann simulation of turbulent flow laden with finite-size
  particles.
\newblock {\em Computers \& Mathematics with Applications}, 65(2):194--210,
  2013.

\bibitem{oka2022attenuation}
Sunao Oka and Susumu Goto.
\newblock Attenuation of turbulence in a periodic cube by finite-size spherical
  solid particles.
\newblock {\em Journal of Fluid Mechanics}, 949:A45, 2022.

\bibitem{peng2023parameterization}
Cheng Peng, Qichao Sun, and Lian-Ping Wang.
\newblock Parameterization of turbulence modulation by finite-size solid
  particles in forced homogeneous isotropic turbulence.
\newblock {\em Journal of Fluid Mechanics}, 963:A6, 2023.

\bibitem{meiburg2010turbidity}
Eckart Meiburg and Ben Kneller.
\newblock Turbidity currents and their deposits.
\newblock {\em Annual Review of Fluid Mechanics}, 42:135--156, 2010.

\bibitem{cantero2009turbidity}
Mariano~I Cantero, S~Balachandar, Alessandro Cantelli, Carlos Pirmez, and Gary
  Parker.
\newblock Turbidity current with a roof: Direct numerical simulation of
  self-stratified turbulent channel flow driven by suspended sediment.
\newblock {\em Journal of Geophysical Research: Oceans}, 114(C3), 2009.

\bibitem{shringarpure2012dynamics}
Mrugesh Shringarpure, Mariano~I Cantero, and S~Balachandar.
\newblock Dynamics of complete turbulence suppression in turbidity currents
  driven by monodisperse suspensions of sediment.
\newblock {\em Journal of Fluid Mechanics}, 712:384--417, 2012.

\bibitem{gore1989effect}
RA~Gore and Clayton~T Crowe.
\newblock Effect of particle size on modulating turbulent intensity.
\newblock {\em International Journal of Multiphase Flow}, 15(2):279--285, 1989.

\bibitem{elghobashi1993two}
S~Elghobashi and GC0782 Truesdell.
\newblock On the two-way interaction between homogeneous turbulence and
  dispersed solid particles. i: Turbulence modification.
\newblock {\em Physics of Fluids A: Fluid Dynamics}, 5(7):1790--1801, 1993.

\bibitem{hetsroni1989particles}
G~Hetsroni.
\newblock Particles-turbulence interaction.
\newblock {\em International Journal of Multiphase Flow}, 15(5):735--746, 1989.

\bibitem{bagchi2004response}
Prosenjit Bagchi and S~Balachandar.
\newblock Response of the wake of an isolated particle to an isotropic
  turbulent flow.
\newblock {\em Journal of Fluid Mechanics}, 518:95--123, 2004.

\bibitem{tanaka2008classification}
Tomohiko Tanaka and John~K Eaton.
\newblock Classification of turbulence modification by dispersed spheres using
  a novel dimensionless number.
\newblock {\em Physical Review Letters}, 101(11):114502, 2008.

\bibitem{luo2016turbulence}
Kun Luo, Mingbo Luo, and Jianren Fan.
\newblock On turbulence modulation by finite-size particles in dilute gas-solid
  internal flows.
\newblock {\em Powder Technology}, 301:1259--1263, 2016.

\bibitem{chouippe2015forcing}
Agathe Chouippe and Markus Uhlmann.
\newblock Forcing homogeneous turbulence in direct numerical simulation of
  particulate flow with interface resolution and gravity.
\newblock {\em Physics of Fluids}, 27(12), 2015.

\bibitem{chouippe2019influence}
Agathe Chouippe and Markus Uhlmann.
\newblock On the influence of forced homogeneous-isotropic turbulence on the
  settling and clustering of finite-size particles.
\newblock {\em Acta Mechanica}, 230:387--412, 2019.

\bibitem{shen2022turbulence}
Jie Shen, Cheng Peng, Jianzhao Wu, Kai~Leong Chong, Zhiming Lu, and Lian-Ping
  Wang.
\newblock Turbulence modulation by finite-size particles of different diameters
  and particle--fluid density ratios in homogeneous isotropic turbulence.
\newblock {\em Journal of Turbulence}, 23(8):433--453, 2022.

\bibitem{cantero2009direct}
Mariano~I Cantero, S~Balachandar, and Gary Parker.
\newblock Direct numerical simulation of stratification effects in a
  sediment-laden turbulent channel flow.
\newblock {\em Journal of Turbulence}, (10):N27, 2009.

\bibitem{salinas2021anatomy}
Jorge~S Salinas, S~Balachandar, M~Shringarpure, J~Fedele, D~Hoyal,
  S~Zu{\~n}iga, and Mariano~Ignacio Cantero.
\newblock Anatomy of subcritical submarine flows with a lutocline and an
  intermediate destruction layer.
\newblock {\em Nature Communications}, 12(1):1649, 2021.

\bibitem{hwang2006homogeneous}
Wontae Hwang and John~K Eaton.
\newblock Homogeneous and isotropic turbulence modulation by small heavy ()
  particles.
\newblock {\em Journal of Fluid Mechanics}, 564:361--393, 2006.

\bibitem{hwang2006turbulence}
Wontae Hwang and John~K Eaton.
\newblock Turbulence attenuation by small particles in the absence of gravity.
\newblock {\em International journal of multiphase flow}, 32(12):1386--1396,
  2006.

\bibitem{peng2019direct}
Cheng Peng, Orlando~M Ayala, and Lian-Ping Wang.
\newblock A direct numerical investigation of two-way interactions in a
  particle-laden turbulent channel flow.
\newblock {\em Journal of Fluid Mechanics}, 875:1096--1144, 2019.

\bibitem{shen2023turbulence}
Jie Shen, Cheng Peng, Jianzhao Wu, Kai~Leong Chong, Zhiming Lu, and Lian-Ping
  Wang.
\newblock Turbulence modulation by finite-size particles of different diameters
  and particle--fluid density ratios in a channel.
\newblock {\em International journal of multiphase flow, revised}, 2023.

\bibitem{bala-book2023}
S~Balachandar.
\newblock {\em Fundamentals of dispersed multiphase flows}.
\newblock Cambridge University Press, 2024.

\bibitem{richardson1954sedimentation}
JF~Richardson and WN~Zaki.
\newblock The sedimentation of a suspension of uniform spheres under conditions
  of viscous flow.
\newblock {\em Chemical Engineering Science}, 3(2):65--73, 1954.

\bibitem{gidaspow1994multiphase}
Dimitri Gidaspow.
\newblock {\em Multiphase flow and fluidization: continuum and kinetic theory
  descriptions}.
\newblock Academic press, 1994.

\bibitem{tenneti2011drag}
S~Tenneti, R~Garg, and S~Subramaniam.
\newblock Drag law for monodisperse gas--solid systems using particle-resolved
  direct numerical simulation of flow past fixed assemblies of spheres.
\newblock {\em International journal of multiphase flow}, 37(9):1072--1092,
  2011.

\bibitem{balachandar2009scaling}
S~Balachandar.
\newblock A scaling analysis for point--particle approaches to turbulent
  multiphase flows.
\newblock {\em International Journal of Multiphase Flow}, 35(9):801--810, 2009.

\bibitem{ling2013scaling}
Y~Ling, M~Parmar, and S~Balachandar.
\newblock A scaling analysis of added-mass and history forces and their
  coupling in dispersed multiphase flows.
\newblock {\em International Journal of Multiphase Flow}, 57:102--114, 2013.

\bibitem{elghobashi1994predicting}
Said Elghobashi.
\newblock On predicting particle-laden turbulent flows.
\newblock {\em Applied scientific research}, 52:309--329, 1994.

\bibitem{yu2021modulation}
Zhaosheng Yu, Yan Xia, Yu~Guo, and Jianzhong Lin.
\newblock Modulation of turbulence intensity by heavy finite-size particles in
  upward channel flow.
\newblock {\em Journal of Fluid Mechanics}, 913:A3, 2021.

\bibitem{wang2000statistical}
Lian-Ping Wang, Anthony~S Wexler, and Yong Zhou.
\newblock Statistical mechanical description and modelling of turbulent
  collision of inertial particles.
\newblock {\em Journal of Fluid Mechanics}, 415:117--153, 2000.

\end{thebibliography}

\end{document}